\documentclass[11pt]{article}
\usepackage{amsmath,amssymb,color,graphics,epsfig}
\usepackage{amsfonts}
\newcommand{\be}{\begin{equation}}
\newcommand{\ee}{\end{equation}}
\newcommand{\bea}{\setlength\arraycolsep{2pt} \begin{eqnarray}}
\newcommand{\eea}{\end{eqnarray}}

\def\ft#1#2{{\textstyle{\frac{\scriptstyle #1}{\scriptstyle #2} } }}
\def\fft#1#2{{\frac{#1}{#2}}}

\def\0{{\sst{(0)}}}
\def\1{{\sst{(1)}}}
\def\2{{\sst{(2)}}}
\def\3{{\sst{(3)}}}
\def\4{{\sst{(4)}}}
\def\5{{\sst{(5)}}}
\def\6{{\sst{(6)}}}
\def\7{{\sst{(7)}}}
\def\8{{\sst{(8)}}}
\def\sst#1{{\scriptscriptstyle #1}}
\def\oneone{\rlap 1\mkern4mu{\rm l}}

\def\R{{\mathbb{R}}}

\begin{document}

\begin{flushright}
%\hfill{KIAS-P12028}
 %\hfill{
%\bf hep-th/yymmnnn}
\end{flushright}

\vspace{25pt}
\begin{center}
{\large {\bf $SL(n,\R)$-Toda Black Holes}}

\vspace{10pt}
H. L\"u and Wei Yang

\vspace{10pt}

{\it Department of Physics, Beijing Normal University,
Beijing 100875, China}

\vspace{40pt}

\underline{ABSTRACT}
\end{center}

We consider $D$-dimensional Einstein gravity coupled to $(n-1)$ $U(1)$ vector fields and $(n-2)$ dilatonic scalars.  We find that for some appropriate exponential dilaton couplings of the field strengths, the equations of motion for the static charged ansatz can be reduced to a set of one-dimensional $SL(n,\R)$ Toda equations.  This allows us to obtain a general class of explicit black holes with mass and $(n-1)$ independent charges.  The near-horizon geometry in the extremal limit is AdS$_2\times S^{D-2}$.   The $n=2$ case gives the Reissner-Nordstr\o m solution, and the $n=3$ example includes the Kaluza-Klein dyon.  We study the global structure and the black hole thermodynamics and obtain the universal entropy product formula.  We also discuss the characteristics of extremal multi-charge black holes that have positive, zero or negative binding energies.

\thispagestyle{empty}

\pagebreak
%\voffset=0pt
%\setcounter{page}{1}

%\tableofcontents
%\addtocontents{toc}{\protect\setcounter{tocdepth}{2}}

%%%%%%%%%%%%%%%%%%%%%%%%%%%%%%%%%%%%%%%%

\newpage
%%%%%%%%%%%%%%%%%%%%%%%%%%%%%%%%%%%%%%%%

\section{Introduction}

Einstein-Maxwell gravity is one of the most natural and simplest examples of gravity coupled to matter. The attempt to unify the two forces in five-dimensional pure gravity turned out to be a failure owing to the unavoidable dilatonic-like massless scalar that couples to the  Kaluza-Klein vector.  The Kaluza-Klein theory nevertheless inspires one to consider a variety of matter fields such as scalars, vectors and higher-rank form fields. In fact the low-energy effective actions of superstrings, namely supergravities, may involve all these fields.  Incidently, only the four-dimensional Einstein-Maxwell theory can be embedded in the string or M-theory.

The string or M-theory provide us a variety of $p$-branes. (See, e.g.~\cite{Duff:1994an}.)  For singly-charged $p$-branes that involve only one field strength, one can always
construct such a solution, extremal \cite{stainless} or non-extremal \cite{Duff:1996hp,Cvetic:1996gq}.  The construction of multi-charge $p$-branes are much subtler and analytical solutions may not exist.  The system can involve multiple field strengths and dilatonic scalars. In this paper, we consider Einstein gravity in $D$ dimensions, coupled with a set of Maxwell fields $A_i$ and dilaton scalars $\vec \phi=(\phi_1,\phi_2,\cdots)$.  The Lagrangian in the form language is given by
\begin{equation}
e^{-1} {\cal L}= R {*\oneone} - \ft12 {*d\vec\phi} \cdot d\vec \phi - \ft12 \sum e^{\vec a_i \cdot \vec \phi}\, {*F_i}\wedge F_i\,,\label{genlag}
\end{equation}
where $F_i=dA_i$ are the $2$-form field strengths and $\vec a_i$  are coupling constant vectors.  It was shown that in maximal supergravities coming from the torus reductions of M-theory, there exists a class of $\ell$-form field strengths whose $\vec a_i$ satisfy the dot product \cite{Lu:1995yn}
\begin{equation}
\vec a_i\cdot \vec a_j = 4\delta_{ij} - \ft{2(\ell-1)(D-\ell-1)}{D-2}\,. \label{aiajliou}
\end{equation}
For these configurations, multi-charge $p$-branes can be constructed \cite{Lu:1995sh}. It turns out that the equations for the $p$-brane ansatz can be reduced to a set of independent Liouville (the $SL(2,\R)$ Toda) equations and hence can be solved more generally \cite{toda}.  Such  cosmological solutions \cite{Lu:1996jk} in the string or M-theory were interpreted as space-like branes \cite{Gutperle:2002ai}. Extremal solutions can be lifted to become intersections of basic objects such as M-branes and D$p$-branes \cite{Tseytlin:1996bh}, whose classifications can be found in \cite{Bergshoeff:1996rn,Lu:1997hb}.

Toda equations are rarer in the $p$-brane constructions in supergravities.  The four-dimensional Kaluza-Klein dyonic black hole (see, e.g.~\cite{Gibbons:1985ac}) is governed by a set of $SL(3,\R)$-Toda equations.  $SL(n,\R)$-Toda instantons or $(D-3)$-branes carrying the axion charges in gravity reduced on $n$-torus were constructed \cite{Lu:1996jr}. Cosmological models with the $E_8$-Toda equation in supergravity were studied in \cite{Lu:1997rd}.

    The goal of this paper is to construct a minimal theory involving
multiple $U(1)$ fields and dilaton scalars that admits multi-charge black holes whose equations of motion are governed by a set of one-dimensional $SL(n,\R)$-Toda equations.  We find that the minimal theory consists of $(n-1)$ vectors and $(n-2)$ scalars, with the $\vec a_i$ satisfying
\begin{eqnarray}
\vec a_i\cdot \vec a_i &=& \ft13(n-2)(n^2+2n+3)\ft{D-3}{D-2}\,,\cr
\vec a_i \cdot  \vec a_{i+1} &=& -\ft16(n^3-n+12)\ft{D-3}{D-2}\,,\cr
\vec  a_i \cdot  \vec a_j &=& -2\ft{D-3}{D-2}\,,\qquad
\hbox{for}\qquad i \ne j-1,\,j,\,j+1\,.\label{aiajtoda}
\end{eqnarray}
Our goal is motivated by the existence of an elegant general solution of the $SL(n,\R)$-Toda equations \cite{todasol} and the previous construction of such $(D-3)$-branes \cite{Lu:1996jr}.  Such theories and general solutions were also studied in
\cite{Ivashchuk:2000yc,Ivashchuk:2001ra}.  The general local solutions contain $2(n-1)$ non-trivial integration constants and they have naked singularity.  In this paper, we obtain proper black holes that run smoothly from the outer horizon to the asymptotic flat space-time.  The black holes have $n$ parameters, associated with the mass and $(n-1)$ independent electric charges.

The paper is organized as follows.  In section 2, we consider the static ansatz of multi-charged black holes for the theory (\ref{genlag}) and derive the equations of motion.  We show that they are a set of Toda-like equations.  These equations can be reduced to a set of independent Liouville equations if the condition (\ref{aiajliou}) is satisfied.  In this paper, we consider instead the conditions (\ref{aiajtoda}).  We start with the simplest case of $SL(2,\R)$ in section 3, corresponding to the Einstein-Maxwell theory.  We then give explicit examples in section 4 and 5 for $SL(3,\R)$ and $SL(4,\R)$ respectively.  We give the general $SL(n,\R)$ black holes in section 6.  While the general solutions with $2(n-1)$ parameters of the $SL(n,\R)$-Toda equations  are elegantly constructed in \cite{todasol}, they are not suitable for describing black holes.  We eliminate $(n-2)$ parameters associated with the scalar deformations that cause naked singularity. The remaining $n$ parameters are associated with the mass and $(n-1)$ independent charges.  The parametrization is presented in a charge-symmetric fashion. In all the cases, we analyse the global structure and obtain all the thermodynamical quantities and verify the first law of thermodynamics.  We also verify that the product of the entropies associated with the outer and inner horizons depend on the (quantized) charges only.  The conclusion and further discussions are given in section 7.

\section{General Formulae}

We begin with a review of general formulae obtained in \cite{stainless,toda}.
Let us consider Einstein gravity in $D$ dimensions, coupled to
a set of $U(1)$ fields $A_i $ and dilaton scalars $\vec \phi=(\phi_1, \phi_2, \cdots)$.  The Lagrangian is given by (\ref{genlag}).  For $k$ $A_i$'s, there can be at most $k$ relevant scalars in (\ref{genlag}). We shall review the equations of motion for the spherically-symmetric configuration that carries electric charges associated with all the $U(1)$ fields.  The ansatz is given by
\begin{equation}
ds_{\sst D}^2=-e^{2A} dt^2 + e^{2B} (dr^2 + r^2 d\Omega_{\sst{D-2}}^2)\,,\qquad
A_i = e^{C_i} dt\,.\label{ansatz1}
\end{equation}
The functions $(A,B,C_i)$, as well as the dilatons $\vec \phi$, depend on the radial coordinate $r$ only.  The equations of motion for the vector fields can be solved easily, giving
\begin{equation}
(e^{C_i})' = \fft{\lambda_i}{r^{D-2}} e^{-\vec a_i \cdot\vec \phi + A - (D-3)B}\,,
\end{equation}
where $\lambda_i$'s are the integration constants, defining the size of the electric charges. The dilaton and Einstein equations of motion become \cite{stainless}
\begin{eqnarray}
&&\vec \phi'' + X' \vec \phi' + \ft{D-2}{r} \vec \phi' = -\ft12 \vec S\,,\cr
&& A'' + X' A' + \ft{D-2}{r} A' = \ft{D-3}{2(D-2)} S\,,\cr
&& B'' + X' B' + \ft{1}{r} X' + \ft{D-2}{r} B' = - \ft{1}{2(D-2)} S\,,\cr
&& X'' -2A'B' + A'^2 - (D-3)B'^2 - \ft{1}{r} X' + \ft12 (\vec\phi')^2 = \ft12 S\,,
\end{eqnarray}
where a prime denotes a derivative with respect to $r$ and $X=A + (D-3) B$.  The functions $\vec S$  and $S$ are
\begin{equation}
\vec S= \fft{1}{r^{2(D-2)}}\sum \vec a_i\lambda_i^2 e^{-\vec a_i\cdot \vec \phi + 2 A-2X}\,,\quad \quad S= \fft{1}{r^{2(D-2)}}\sum \lambda_i^2 e^{-\vec a_i\cdot \vec \phi + 2 A-2X}\,.
\end{equation}
In \cite{stainless} only extremal $p$-branes with the condition $X=0$ were considered.  As was shown in \cite{toda}, if one defines a new coordinate $\rho=r^{3-D}$.  The function $X$ can be easily solved, giving
\begin{equation}
X_{,\rho\rho} + X_{,\rho}^2 - \rho^{-1} X_{,\rho}=0\,,\qquad\Longrightarrow \qquad e^X =1 - k^2 \rho^2\,.
\end{equation}
It is now convenient to define a new coordinate $\xi$ by
\begin{equation}
d\xi = e^{-X} d\rho\,,\qquad \hbox{\it i.e.}\qquad k\rho=\tanh(k\xi)\,.
\end{equation}
The equations of motion can be reduced to some Toda-like equations, namely
\cite{toda}
\begin{eqnarray}
\ddot {\vec \phi} &=& -\ft{1}{2 (D-3)^2} \sum \vec a_I \lambda_i^2 e^{-\vec a_i\cdot \vec \phi + 2 A}\,,\cr
\ddot A &=& \ft{1}{2(D-2)(D-3)} \sum \lambda_i^2 e^{-\vec a_i\cdot \vec \phi + 2 A}\,,\label{todalike}
\end{eqnarray}
together with the first-order Hamiltonian constraint
\begin{equation}
(D-3)(D-2) \dot A^2 + \ft12 (D-3)^2 \dot{\vec\phi} \cdot \dot{\vec\phi} -\ft12 \sum\lambda_i^2  e^{-\vec a_i\cdot \vec \phi + 2 A}=4(D-2)(D-3) k^2\,.\label{todalikefo}
\end{equation}
Here a dot denotes a derivative with respect to $\xi$.

    Thus the structure of the equations are fully governed by the
dilaton coupling constant vectors $\vec a_i$, in particular their dot products $\vec a_i \cdot \vec a_j$.  In general,
the equations (\ref{todalike}) and (\ref{todalikefo}) cannot be solved analytically.  It was shown in \cite{Lu:1995yn,toda} that if the dot products (\ref{aiajliou}) are satisfied, the equations can be reduced to a set of independent Liouville equations and hence can be fully solved.
In the follow sections, we shall consider some special choice of dilaton vectors such that the equations (\ref{todalike}) are rendered to be the one-dimensional $SL(n,\R)$-Toda equations and hence become completely solvable.

\section{$SL(2,\R)$-Toda (Liouville) black hole}

The simplest case is the $SL(2,\R)$, which gives the Liouville equation. The minimal theory whose black hole equations of motion can give rise to a Liouville equation is the Einstein-Maxwell theory
\begin{equation}
e^{-1}{\cal L}= R - \ft14 F^2\,,
\end{equation}
where $F=dA$. (Note that there should be no confusion between the 1-form $A$ and the same-named function in the metric ansatz (\ref{ansatz1}).) Following the discussion in the previous section,  the equations of motion can be reduced to
\begin{equation}
\ddot A = \ft{\lambda^2}{2(D-2)(D-3)} e^{2A}\,.
\end{equation}
together with the first-order constraint
\begin{equation}
(D-3)(D-2) \dot A^2 - \ft12 \lambda^2 e^{2A} = 4(D-2)(D-3) k^2\,.
\end{equation}
Thus the solution is given by
\begin{equation}
e^{-A} = \tilde \lambda \sinh(2k\xi + \gamma)
\,,
\end{equation}
where $\tilde\lambda = \lambda/(2\sqrt{2(D-2)(D-3)}\,k)$.
The function $e^C$ associated with vector potential is given by
\begin{equation}
e^{C} = \sqrt{\ft{2(D-2)}{(D-3)\tilde\lambda^2}}\, \coth(2k\xi + \gamma)\,.
\end{equation}
Introducing a new radial coordinate $r$, (which is different from the $r$ in (\ref{ansatz1}),) defined by
\begin{equation}
e^{2k \xi} = \fft{1}{\sqrt f}\,,\qquad f=1 - \fft{m}{r^{D-3}}\,,\qquad m=4k\,,\label{xitor}
\end{equation}
we find that the solution is now given by
\begin{eqnarray}
ds^2 &=&-H^{-2} f dt^2 + H^{\fft{2}{D-3}} (f^{-1} dr^2 + r^2 d\Omega_{D-2}^2)\,,\cr
A &=& \sqrt{\ft{2(D-2)}{D-3}} \coth\delta\, H^{-1} dt\,,\qquad
H=1 + \fft{m \sinh^2\delta}{r^{D-3}}\,,
\end{eqnarray}
where we have renamed $\tilde\lambda = \sinh 2\delta$.  This is precisely the Reissner-Nordstr\o m (RN) black hole written in the $p$-brane coordinates. This solution has been well studied and we shall not consider it further.

It is worth pointing out however that the coordinate transformation (\ref{xitor}) is universal in all of the subsequent discussions. The range from the horizon at $r_0=m^{1/(D-3)}$ to the asymptotic $r=\infty$ is mapped to that from $\xi=\infty$ to $\xi=0$.

\section{$SL(3,\R)$-Toda black hole}

In the previous section, we demonstrate that the equations of motion for the RN black hole can be reduced to a Liouville equation.  We now consider Einstein gravity coupled to two $U(1)$ vectors $(A_1,A_2)$ and one dilaton $\phi$.  The general Lagrangian is given by
\begin{equation}
e^{-1} {\cal L} = R - \ft12 (\partial\phi)^2 - \ft14 e^{a_1\phi} F_1^2 - \ft14 e^{a_2\phi} F_2^2\,.
\end{equation}
The equations for two-charged black holes were studied in \cite{HL}.  It can be reduced to Toda-like equations and cannot be solved in general.  When $a_1 a_2=-2(D-3)/(D-2)$, the Toda-like equations can be reduced to two independent Liouville equations and hence can be solved completely.  This gives rise to black holes with two independent charges.  In this paper, we consider an alternative choice.  We find that when $(a_1,a_2)$ are
\begin{equation}
a_1=-a_2 = a \equiv \sqrt{\ft{6(D-3)}{D-2}}\,,
\end{equation}
the Toda-like equations become the true Toda equation of $SL(3,\R)$, which allows us to obtain the most general analytical solution.  To see this in some detail, let us first make the following field redefinition
\begin{equation}
\phi = - \ft{a(D-2)}{4(D-3)} (q_1 - q_2) + \ft{1}{a} \log \ft{\lambda_1}{\lambda_2}\,,\qquad
A=\ft14 (q_1 + q_2) +\ft12 \log\ft{(D-2)(D-3)}{2\lambda_1\lambda_2}\,.
\end{equation}
We find that the equations of motion are reduced to the one-dimensional $SL(3,\R)$-Toda equations:
\begin{equation}
\ddot q_1 = e^{2q_1 - q_2}\,,\qquad \ddot q_2 = e^{-q_1 + 2 q_2}\,,\label{sl2reom}
\end{equation}
together with the first-order Hamiltonian constraint
\begin{equation}
\ft1{16} (\dot q_1 + \dot q_2)^2 + \ft{3}{16} (\dot q_1 - \dot q_2)^2 -
\ft14 (\ddot q_1 + \ddot q_2) = 4k^2\,.\label{sl2rfo}
\end{equation}
Note that to be qualified as the {\it first-order} Hamiltonian constraint, we should really replace $\ddot q_i$  by those in (\ref{sl2reom}). The current expression however has its charm of being shorter.
Analogous equations were obtained for dyonic $p$-branes in \cite{toda}.
The $SL(3,\R)$ equation can be completely solved.  The general solution contains four non-trivial integration constants and it has a naked curvature singularity.  For the solution to be well-defined from the horizon at $\xi=\infty$ to the asymptotic infinity $\xi=0$, one parameter has to eliminated. The procedure is outlined in \cite{toda} for the construction of dyonic $p$-branes.  In order to avoid excessive repetition, we shall simply present the final answer of the black hole solution
\begin{eqnarray}
ds^2&=& -(H_1H_2)^{-\fft12} f\,  dt^2 + (H_1H_2)^{\fft{1}{2(D-3)}}\Big(f^{-1} dr^2 + r^2 d\Omega_{D-2}^2\Big)\,,\cr
\phi &=& \ft12 \sqrt{\ft{3(D-2)}{2(D-3)}}\, \log(\fft{H_1}{H_2})\,,\qquad f=1 - \fft{m}{r^{D-3}}\,,\cr
A_1&=&\sqrt{\ft{D-2}{D-3}}\,\fft{1-\beta_1 f}{\sqrt{\beta_1\gamma_2}\,H_1}\,dt\,,\qquad
A_2=\sqrt{\ft{D-2}{D-3}}\,\fft{1-\beta_2 f}{\sqrt{\beta_2 \gamma_1} \,H_2}\,dt\,,\cr
H_1&=&\gamma_1^{-1} (1-2\beta_1 f + \beta_1\beta_2 f^2)\,,\qquad
H_2=\gamma_2^{-1}(1 - 2\beta_2 f + \beta_1\beta_2 f^2)\,,\cr
\gamma_1&=& 1- 2\beta_1 + \beta_1\beta_2\,,\qquad \gamma_2 = 1-2\beta_2 + \beta_1\beta_2\,.
\label{sl3rsolution1}
\end{eqnarray}
The solution is written in the symmetric form in terms of the charge parameters $\beta_i$.  When $\beta_1$ or $\beta_2$ is set to zero, the solution is reduced to be only singly charged.  If $\beta_1=\beta_2$, the solution becomes the RN black hole.  for the general solution, we require that $\beta_i>0$ and $\gamma_i\ge 0$ for it to be well behaved from the horizon to asymptotic infinity.  This implies that we cannot impose the range for $\beta_i$ independently.  We may chose a different parametrization
\begin{equation}
\beta_i=\lambda_j^{-1} (1 - \sqrt{1-\lambda_1 \lambda_2})\,,\qquad i \ne j\,,
\end{equation}
the parameters $\lambda_i$ can then independently take the values lying in the range $[0,1]$.

   The $SL(3,\R)$ black hole approaches the Minkowski space-time in the
$r\rightarrow \infty$ asymptotic region.  From the falloffs, we can read off the ADM mass:
\begin{equation}
M=\fft{(D-2)\Omega\, (1-\beta_1)(1-\beta_2)(1-\beta_1\beta_2)\, m}{16\pi \gamma_1\gamma_2}
\end{equation}
where $\Omega$ denotes the volume of the unit round $(D-2)$-sphere.
The electric charges are given by
\begin{equation}
Q_1=\fft{\Omega\sqrt{(D-2)(D-3)\beta_1\gamma_2}\,m}{16\,\pi \gamma_1}\,,\qquad
Q_2=\fft{\Omega\sqrt{(D-2)(D-3)\beta_2\gamma_1}\,m}{16\,\pi \gamma_2}\,.
\end{equation}
The outer horizon is located at $r_0=m^{1/(D-3)}$. The temperature and entropy can be calculated using the standard method and they are
\begin{equation}
T=\ft{D-3}{4\pi r_0}\,(\gamma_1\gamma_2)^{\fft{D-2}{4(D-3)}}\,,\qquad S=\ft14\Omega \,(
\ft{m^4}{\gamma_1\gamma_2})^{\fft{D-2}{4(D-3)}}\,.
\end{equation}
Electric potential differences between the horizon and the asymptotic infinity for the two vectors are given by
\begin{equation}
\Phi_1 = (1-\beta_2) \sqrt{\ft{(D-2)\beta_1}{(D-3)\gamma_2}}
\,,\qquad
\Phi_2 = (1-\beta_1) \sqrt{\ft{(D-2)\beta_2}{(D-3)\gamma_1}}\,.
\end{equation}
we find that the first law of thermodynamics
\begin{equation}
dM = T dS + \Phi_1 dQ_1 + \Phi_2 dQ_2
\end{equation}
is satisfied.

The extremal solution can be obtained by taking the appropriate $\gamma_i=0$ limit, corresponding to $\beta_i=1$, together with $m=0$.  Making the reparametrization
\begin{eqnarray}
\beta_1 &=& 1 - q_1^{-\fft23}q_2^{-\fft23} (q_1^{\fft23} + q_2^{\fft23})^{\fft12}\,m + q_1^{-\fft23} q_2^{-\fft43}\,m^2\,,\cr
\beta_2 &=& 1 - q_1^{-\fft23}q_2^{-\fft23} (q_1^{\fft23}+
q_2^{\fft23})^{\fft12}\, m + q_1^{-\fft43} q_2^{-\fft23}\,m^2\,,
\end{eqnarray}
and then letting $m\rightarrow 0$, we find that $f=1$ and
\begin{eqnarray}
H_1 &=&  1 + q_1^{\fft23}(q_1^{\fft23} + q_2^{\fft23})^{\fft12}
\, \rho + \ft12 q_1^{\fft43} q_2^{\fft23}\,\rho^2\,,\cr
H_2 &=&  1 + q_2^{\fft23}(q_1^{\fft23} + q_2^{\fft23})^{\fft12}
\, \rho + \ft12 q_1^{\fft23} q_2^{\fft43}\,\rho^2\,,\cr
A_i&=&\sqrt{\ft{D-2}{2(D-3)}} \Big( (q_1^{\fft23} + q_2^{\fft23})^{\fft12} + q_1^{\fft23} q_2^{\fft23}\,\rho\Big) q_i^{-\fft13} H_i^{-1}\,dt\,,
\end{eqnarray}
where $\rho=1/r^{D-3}$.  It is easy to verify that the mass in the extremal limit is given by
\begin{equation}
M_{\rm ext} = \sqrt{\ft{D-2}{2(D-3)}} \Big(Q_1^{\fft23} + Q_2^{\fft23}\Big)^{\fft32}\,,
\end{equation}
where $Q_i= (q_i/16)\sqrt{(D-2)(D-3)/2}$.

The horizon for the extremal limit is located at $r=0$, and the horizon geometry is an AdS$_2\times S^{D-2}$ with the residual entropy given by
\begin{equation}
S_{\rm ext} = \Big( \fft{2^{\fft{4(D-1)}{D-2}} \pi^2 Q_1 Q_2}{(D-2)(D-3) \Omega^{\fft2{D-2}}}\Big)^{\fft{D-2}{2(D-3)}}\,.
\end{equation}
The electric potentials are given by
\begin{equation}
\Big(\Phi_1^{\rm ext}\,,\,\,\Phi_2^{\rm ext}\Big) = \sqrt{\ft{D-2}{2(D-3)}}\, (Q_1^{\fft23} +
Q_2^{\fft23})^{\fft12}\, \Big(Q_1^{-\fft13}\,,\,\, Q_2^{-\fft13}\Big)\,.
\end{equation}
The mass and charges in the extremal limit continues to satisfy the ``first law'' of thermodynamics at the zero temperature
\begin{equation}
dM_{\rm ext}=\Phi_1^{\rm ext} dQ_1 + \Phi_2^{\rm ext} dQ_2\,.
\end{equation}

When the solution is non-extremal, there is an inner horizon located at $r=0$, and its entropy is given by
\begin{equation}
S_- = \ft14 \Omega\, \Big(\fft{m^4 \beta_1^2\beta_2^2}{\gamma_1\gamma_2}\Big)^{\fft{D-2}{4(D-3)}}\,.
\end{equation}
If we rename the entropy on the outer horizon as $S_+$, we have the product of the two entropies
\begin{equation}
S_+ S_- =S_{\rm ext}^2\,.
\end{equation}
The product depends on the charges only and hence is quantized.

   Note that in four dimensions, the two charges can arise as the electric and magnetic
charges of the same field strength, corresponding to the well-known Kaluza-Klein dyon, (see, e.g.~\cite{Gibbons:1985ac}.)

\section{$SL(4,\R)$-Toda black hole}

The minimum theory that gives rise to an $SL(4,\R)$-Toda black hole involves the metric, three $U(1)$ vectors $(A_1,A_2,A_3)$ and two dilatonic scalars $\vec \phi=(\phi_1,\phi_2)$, with the Lagrangian (\ref{genlag}), with $i=1,2,3$. The $\vec a_i$'s are two-component constant vectors.  It can be shown that when $\vec a_i$'s satisfy the condition
\begin{equation}
\vec a_i\cdot \vec a_j = - \ft{2(D-3)}{D-2}\,,\qquad \hbox{for}\qquad i\ne j\,,\label{aiajliouville}
\end{equation}
the equations can be reduced to three independent Liouville equations.  In order to construct the $SL(4,\R)$-Toda black hole, we find that $\vec a_i$ must satisfy
\begin{eqnarray}
&&\vec a_1 \cdot \vec a_1=\vec a_2 \cdot \vec a_2=\vec a_3 \cdot \vec a_3=
\ft{18(D-3)}{D-2}\,,\cr
&& \vec a_1\cdot \vec a_2=\vec a_2 \cdot \vec a_3 = -\ft{12(D-3)}{D-2}\,,\qquad
\vec a_1\cdot \vec a_3 = -\ft{2(D-3)}{D-2}\,.\label{sl4raiaj}
\end{eqnarray}
Note that the three $\vec a_i$ vectors are spanned in two-dimensional space and cannot be linearly dependent.  It follows from (\ref{sl4raiaj}) that we have
\begin{equation}
3 \vec a_1 + 4 \vec a_2 + 3 \vec a_3=0\,.
\end{equation}
The solutions to the conditions (\ref{sl4raiaj}) are not unique; an orthonormal rotation between $\phi_1$ and $\phi_2$ gives an equivalent choice for the vectors $\vec a_i$. Here we present one explicit example
\begin{equation}
\vec a_1 = \sqrt{\ft{D-3}{D-2}}(3\sqrt{2},\, 0)\,,\quad
\vec a_2 = \sqrt{\ft{D-3}{D-2}}(-2\sqrt{2},\, \sqrt{10})\,,\quad
\vec a_3 = \sqrt{\ft{D-3}{D-2}}(-\ft13\sqrt{2},\, -\ft43\sqrt{10})\,.
\end{equation}
In order to see that the the resulting equations of motion can be cast into the
Toda equations, we make the following field redefinitions
\begin{eqnarray}
e^{-\vec a_1 \cdot \vec\phi + 2A} &=& \ft15 (D-2)(D-3)\lambda_1^{-2}\, e^{2q_1 - q_2}\,,\cr
e^{-\vec a_2 \cdot \vec \phi + 2A} &=& \ft15(D-2)(D-3)\lambda_2^{-2}\, e^{-q_1+2q_2 - q_3}\,,\cr
e^{-\vec a_3 \cdot \vec \phi + 2A} &=& \ft15 (D-2)(D-3)\lambda_3^{-2}\, e^{-q_2+2q_3}\,.
\end{eqnarray}
In this field redefinition, the fields $(\phi_1,\phi_2)$ and the function $A$ are mapped to functions $(q_1,q_2,q_3)$.
This redefinition is only possible provided that the dot products (\ref{sl4raiaj}) are satisfied.  In fact, that is how we determined these dot products in (\ref{sl4raiaj}).  The equations of motion now become
\begin{equation}
\ddot q_1 = e^{2q_1 - q_2}\,,\qquad
\ddot q_2 = e^{-q_1+2q_2 - q_3}\,,\qquad
\ddot q_3 =e^{-q_2+2q_3}\,.\label{sl4reom}
\end{equation}
These are precisely the one-dimensional $SL(4,\R)$-Toda equations.  The first-order Hamiltonian constraint is given by
\begin{equation}
\ft1{100} (\dot q_1 + \dot q_2 + \dot q_3)^2 +
\ft1{900} (9\dot q_1 - 6 \dot q_2 - \dot q_3)^2 +
\ft{1}{180} (3\dot q_2 - 4\dot q_3)^2 - \ft{1}{10} (\ddot q_1 + \ddot q_2 + \ddot q_3) = 4k^2\,.\label{sl4rhamil}
\end{equation}
Again, the two-derivative terms in the above should be replaced by the exponential terms {\it via} the $SL(4,\R)$ Toda equation (\ref{sl4reom}).
The most general solution is given by \cite{todasol}
\begin{equation}
e^{-q_1} = \sum_{i=1}^4 c_i e^{\mu_i \xi}\,,\qquad e^{-q_2} = \sum_{i<j}^4 b_{ij} e^{(\mu_i + \mu_j)\xi}\,,\qquad e^{-q_3} = \sum_{i=1}^4 d_i e^{-\mu_i \xi}\,,
\end{equation}
where
\begin{eqnarray}
&&b_{ij} = - c_i c_j (\mu_i-\mu_j)^2\,,\qquad
d_i^{-1} = -c_i  \prod_{j\ne i}^4 (\mu_i-\mu_j)^2\,,\cr
&&\mu_1 + \mu_2 + \mu_3 + \mu_4=0\,,\qquad
c_1c_2c_3c_4\prod_{i<j}^4 (\mu_i-\mu_j)^2 =1\,.
\end{eqnarray}
The Hamiltonian constraint (\ref{sl4rhamil}) implies that
\begin{equation}
\mu_1^2 + \mu_2^2 + \mu_3^2 + \mu_1\mu_2 + \mu_1 \mu_3 + \mu_2\mu_3=40k^2\,.
\end{equation}
The general solution contains a total of six parameters and it has naked singularity.  In our approach we make an implicit assumption that the black hole horizon is located at $\xi=\infty$ whilst the asymptotic infinity is at $\xi=0$.  In order for the solution to be absent from naked singularity, we require that $\vec \phi$ is non-divergent at both $\xi=0$ and $\xi=\infty$.  This implies that $e^{q_1-q_3}$ and $e^{4q_1-3q_2}$ are non-divergent at both $\xi=0$ and $\xi=\infty$.  We find that the solution to this requirement is that
\begin{equation}
(\mu_1,\mu_2,\mu_3) = (6k, 2k, -2k)\,,\label{mu123res}
\end{equation}
and hence $\mu_4=-6k$.  With these constraints, our solution now contains four independent parameters. Making the coordinate transformation (\ref{xitor}) and the reparametrization
\begin{equation}
(c_1,c_2,c_3,c_4) = \ft1{384 k^3} ( (\beta_1^3\beta_2^2\beta_3)^{-\fft14}\,,
3(\beta_1\beta_2^{-2}\beta_3^{-1})^{\fft14}\,, 3(\beta_1\beta_2^2\beta_3^{-1})^{\fft14}\,,
(\beta_1\beta_2^2\beta_3^3)^{\fft14})\,,
\end{equation}
we find that the $SL(4,\R)$-Toda black hole is given by
\begin{eqnarray}
ds^2 &=& - (H_1 H_2 H_3)^{-\fft15} f dt^2 + (H_1 H_2 H_3)^{\fft{1}{5(D-3)}} (f^{-1} dt^2 + r^2 d\Omega_{D-2}^2)\,,\cr
e^{\vec a_1 \cdot \vec \phi} &=& \fft{H_1^2}{H_2 (H_1 H_2H_3)^{\fft15}}\,,\qquad e^{\vec a_2\cdot \vec \phi} = \fft{H_2^2}{H_1 H_3 (H_1 H_2H_3)^{\fft15}}\,,\cr
e^{\vec a_3\cdot \vec \phi} &=& \fft{H_3^2}{H_2 (H_1 H_2H_3)^{\fft15}}\,,\qquad f=1-\fft{m}{r^{D-3}}\,,\cr
A_1 &=& \sqrt{\ft{3(D-2)}{5(D-3)}}\, \fft{1-2 \beta_1 f + \beta_1\beta_2 f^2}{\sqrt{\beta_1\gamma_2}\, H_1}\,dt\cr
A_2 &=& \sqrt{\ft{4(D-2)}{5(D-3)}}\,\fft{1-3\beta_2 f + \ft32 \beta_2(\beta_1 + \beta_2) f^2 -\beta_1\beta_2\beta_3 f^3}{\sqrt{\beta_2\gamma_1\gamma_3}\, H_2}\,dt\,,\cr
A_3 &=&  \sqrt{\ft{3(D-2)}{5(D-3)}}\, \fft{1-2\beta_3 f+ \beta_2\beta_3 f^2}{\sqrt{\beta_3\gamma_2}\,H_3}\,dt\,,
\end{eqnarray}
where
\begin{eqnarray}
H_1 &=& \gamma_1^{-1} (1 - 3 \beta_1 f + 3 \beta_1 \beta_2 f^2 - \beta_1 \beta_2 \beta_3 f^3)\,,\cr
H_2 &=& \gamma_2^{-1} (1 - 4 \beta_2 f + 3 \beta_2 (\beta_1 + \beta_3) f^2 - 4 \beta_1 \beta_2 \beta_3 f^3 +
 \beta_1 \beta_2^2 \beta_3 f^4)\,,\cr
H_3&=& \gamma_3^{-1} (1 - 3 \beta_3 f + 3 \beta_2 \beta_3 f^2 - \beta_1 \beta_2 \beta_3 f^3)\,,
\end{eqnarray}
with
\begin{eqnarray}
\gamma_1 &=&1 - 3 \beta_1 + 3 \beta_1 \beta_2 - \beta_1 \beta_2 \beta_3\,,\cr
\gamma_2 &=&1 - 4 \beta_2 + 3 \beta_2(\beta_1 + \beta_3) - 4 \beta_1 \beta_2 \beta_3 + \beta_1 \beta_2^2 \beta_3\,, \cr
\gamma_3 &=&1 - 3 \beta_3 + 3 \beta_2 \beta_3 -
 \beta_1 \beta_2 \beta_3\,.
 \end{eqnarray}
We can turn off any charge by setting the associated $\beta_i$ to zero.  When all the $\beta_i$ are equal, the solution reduces to the RN black hole.  The thermodynamical quantities can be easily obtained.  The mass is somewhat complicated, given by
\begin{eqnarray}
M&=&\fft{(D-2)\Omega\,m}{16\pi} \Big(1 + \fft{3\beta_1}{5\gamma_1}(1-2\beta_2+\beta_2\beta_3)\cr
&&+ \fft{4\beta_2}{5\gamma_2}(1 - \ft32(\beta_1+\beta_2) + 3\beta_1\beta_3 - \beta_1\beta_2\beta_3)+ \fft{3\beta_3}{5\gamma_3} (1 - 2 \beta_2 + \beta_1 \beta_2)\Big)\,.
\end{eqnarray}
The electric charges are given by
\begin{eqnarray}
Q_1&=&\fft{\Omega\,m}{16\pi \gamma_1}\sqrt{\ft35(D-2)(D-3)\beta_1\gamma_2}\,,\cr
Q_2&=&\fft{\Omega\,m}{16\pi \gamma_2}\sqrt{\ft45(D-2)(D-3)\beta_2\gamma_1\gamma_3}\,,\cr
Q_3&=&\fft{\Omega\,m}{16\pi \gamma_3}\sqrt{\ft35(D-2)(D-3)\beta_3\gamma_2}\,.
\end{eqnarray}
Their thermodynamical conjugate electric potentials are
\begin{eqnarray}
\Phi_1&=&\sqrt{\ft{3(D-2)\beta_1}{(D-5)\gamma_2}}\, (1-2\beta_2 + \beta_2\beta_3)\,,\cr
\Phi_2&=&\sqrt{\ft{4(D-2)\beta_2}{(D-5)\gamma_1\gamma_3}}\, (1-
\ft32(\beta_1 + \beta_3) + 3\beta_1\beta_3 -\beta_1\beta_2\beta_3)\,,\cr
\Phi_3&=&\sqrt{\ft{3(D-2)\beta_3}{(D-5)\gamma_2}}\, (1-2\beta_2 + \beta_1\beta_2)\,,
\end{eqnarray}
The temperature and entropies can be obtained using the standard technique, given by
\begin{equation}
T=\ft{D-3}{4\pi r_0} (\gamma_1 \gamma_2 \gamma_3)^{\fft{D-2}{10(D-3)}}\,, \qquad
S=\ft14 \Omega \Big(\ft{m^{10}}{\gamma_1 \gamma_2 \gamma_3} \Big)^{ \fft{D-2}{10(D-3)}}\,.
\end{equation}
It is straightforward to verify the first law of thermodynamics
\begin{equation}
dM=T dS + \Phi_1 dQ_1 + \Phi_2 dQ_2 + \Phi_3 dQ_3\,.
\end{equation}

The extremal limit corresponds to setting $m=0$ while keeping all the charges $Q_i$ non-vanishing.  This can be achieved by letting $\beta_i\rightarrow 1$.  To be specific, let us reparametrize $\beta_i$ in terms of constants $(a,b,c)$:
\begin{equation}
\beta_1 = 1 - a\, m +  a^2 b\, m^2\,,\qquad
\beta_2 = 1 - a\, m - \ft12 a^3\, (c-1)\, m^3\,,\qquad
\beta_3 = 1 - a\, m -  a^2 b\, m^2\,,
\end{equation}
and then let $m\rightarrow 0$. In this limit, we find
\begin{equation}
q_1 = \ft{\sqrt{3(3 + 3 b^2 -2 c)}}{a(c-3b)}\,,\qquad
q_2 = \ft{2\sqrt{c^2 - 9 b^2}}{a(3 + 3 b^2 - 2c)}\,,\qquad
q_3 = \ft{\sqrt{3(3 + 3b^2 - 2 c)}}{a(c+3b)}\,,
\end{equation}
where the electric charges are now given by
\begin{equation}
(Q_1,Q_2,Q_3) = \ft{\Omega}{16\pi}\sqrt{\ft15(D-2)(D-3)}\,(q_1,\, q_2,\, q_3)\,.
\end{equation}
The functions $H_i$ in the extremal limit is given by
\begin{eqnarray}
H_1 &=& 1 + \ft{3(1-b)}{a(c-3b)}\rho + \ft{3}{a^2(c-3b)} \rho^2 + \ft1{a^3(c-3b)}\rho^3\,,\cr
H_2 &=& 1 +\ft{2 (3 - c)}{a (3 + 3 b^2 - 2 c)} \rho+ \ft{6}{
 a^2 (3 + 3 b^2 - 2 c)} \rho^2+ \ft{4}{a^3 (3 + 3 b^2 - 2 c)}\rho^3 + \ft1{
 a^4 (3 + 3 b^2 - 2 c)}\rho^4\,,\cr
H_3 &=& 1 + \ft{3(1+b)}{a(c+3b)}\rho + \ft{3}{a^2(c+3b)} \rho^2 + \ft1{a^3(c+3b)}\rho^3\,,
\end{eqnarray}
where $\rho=1/r^{D-3}$. The mass of the extremal black hole is given by
\begin{equation}
M=\ft{(D-2)\Omega(-54 b^2 - 27 b^4 + 9 c + 36 b^2 c - 3 c^2 - c^3)}{
 40\pi a (3 + 3 b^2 - 2 c) (c^2 - 9b^2) }
\end{equation}
As was shown in \cite{toda}, the mass is a sixth-order polynomial in terms of the three independent charges.  The expression becomes simpler if we let $Q_3=Q_1$, which implies that
\begin{equation}
a=\ft{\sqrt{3-2c}}{c q_1}\,,\qquad b=0\,,\qquad c^2 q_1 = (3 - 2 c)^{\fft32} q_2\,.
\end{equation}
Let $M=(D-2)\Omega\widetilde M/(80 \pi)$, we find in this case that
\begin{equation}
\widetilde M^4 +2q_2 \widetilde M^3 - 8 q_1^2 \widetilde M^2 + 2q_2 (28 q_1^2 - q_2^2) \widetilde M + 16 q_1^4 - 44 q_1^2 q_2^2 - q_2^4=0\,.
\end{equation}
Thus we see that $\widetilde M=2q_1$ for $q_2=0$ and $\widetilde M=q_2$ for $q_1=0$.  For non-vanishing $q_i$, $\widetilde M$ is bigger than the sum of individual $q_i$, implying that the bound state has negative binding energy, as in the case of $SL(3,\R)$-Toda black hole discussed in the previous section.

    The extremal solution is much less elegant compared to its non-extremal counterpart.
In \cite{toda}, a different parametrization is given, with the three independent parameters all defined in $H_1$, namely
\begin{equation}
H_1 = 1 + a \rho + b \rho^2 + c\rho^3\,.
\end{equation}
The remaining $H_i$'s are then fully determined by the Toda equations and the vanishing of the first-order Hamiltonian constraint. It would be of interest to obtain a charge-symmetric parametrization.

\section{$SL(n,\R)$-Toda black hole}

   We are now in the position to present the general $SL(n,\R)$-Toda black holes.  The
minimal matter consists of $(n-1)$ $U(1)$ fields and $(n-2)$ dilatonic scalars.  The Lagrangian takes the form (\ref{genlag}), with the dilaton coupling constant vectors $\vec a_i$  satisfying the dot products (\ref{aiajtoda}).
The $a_i$'s are not linearly independent, but satisfy the linear equation
\begin{equation}
\sum_{i=1}^{n-1} i (n-i) \vec a_i = 0\,.
\end{equation}
With these, we find that it is consistent to make the field redefinitions
\begin{eqnarray}
e^{-\vec a_1 \cdot \vec\phi + 2A} &=& \ft{12 (D-2)(D-3)}{n(n^2-1)} \lambda_1^{-2}\, e^{2q_1 - q_2}\,,\cr
e^{-\vec a_2 \cdot \vec \phi + 2A} &=& \ft{12 (D-2)(D-3)}{n(n^2-1)}\lambda_2^{-2}\, e^{-q_1+2q_2 - q_3}\,,\cr
e^{-\vec a_3 \cdot \vec \phi + 2A} &=& \ft{12 (D-2)(D-3)}{n(n^2-1)}\lambda_3^{-2}\, e^{-q_2+2q_3-q_4}\,,\cr
&\vdots&\cr
e^{-\vec a_{n-1} \cdot \vec\phi + 2A} &=& \ft{12(D-2)(D-3)}{n(n^2-1)}\lambda_{n-1}^{-2}\, e^{-q_{n-2} + 2 q_{n-1}}\,.
\end{eqnarray}
The resulting equations of motion become the one-dimensional $SL(n,\R)$-Toda equations
\begin{eqnarray}
\ddot q_1 &=& e^{2q_1 - q_2}\,,\cr
\ddot q_2 &=& e^{-q_1 + 2 q_2 - q_3}\,,\cr
\ddot q_3 &=& e^{-q_2 + 2q_3 - q_4}\,,\cr
&\vdots&\cr
\ddot q_{n-1} &=& e^{-q_{n-2} + 2 q_{n -1}}\,,\label{slnrtodaeom}
\end{eqnarray}
together with the first-order Hamiltonian constraint:
\begin{equation}
\sum_{i=1}^{n-1} (\dot q_i^2-\ddot q_i) -\sum_{i<j} \dot q_i \dot q_j = \ft23 n(n^2-1) k^2\,.
\end{equation}
Note that here each $\ddot q_i$ should be replaced by the corresponding the right-hand side of (\ref{slnrtodaeom}). An elegant and general solution for this system, but with a minus sign in all the right sides of the equations, was obtained in \cite{todasol}.  Taking some appropriate constant shift on each $q_i$ to compensate this minus sign, we find
\begin{eqnarray}
&&e^{-q_i} = (-1)^{\fft12i(i-1)} \sum_{k_1<\cdots<k_i}^n c_{k_1} \cdots c_{k_i} \Delta^2(k_1,\ldots,k_i) e^{(\mu_{k_1} +\cdots + \mu_{k_i})\xi}\,,\cr
&&\Delta^2(k_1,\ldots, k_i) \equiv \prod_{k_i<k_j} (\mu_{k_i} - \mu_{k_j})^2\,,
\end{eqnarray}
where the constants $c_k$ and $\mu_k$ are arbitrary constants satisfying
\begin{equation}
\prod_{i=1}^n c_i = (-1)^{\fft12n(n-1)} \Delta^{-2}(1,2,\ldots,n)\,,\qquad
\sum_{i=1}^n \mu_i=0\,.
\end{equation}
The Hamiltonian constraint is given by
\begin{equation}
\sum_{i}^n \mu_i^2 = \ft23 n(n^2-1) k^2 \equiv \ft1{24} n(n^2-1) m^2\,.
\end{equation}
Analogous location solutions were also given in \cite{Ivashchuk:2000yc,Ivashchuk:2001ra}; however, they do not describe a black hole.  In fact the general solution contains $2(n-1)$ non-trivial integration constants, and $(n-2)$ of them are associated with the $(n-2)$ scalar charges and should be eliminated in order to have the flat asymptotics and avoid naked singularity. We find that this can be achieved by letting
\begin{equation}
\mu_i =\ft12 m (n +1 - 2i)\,,\qquad i=1,2,\cdots,n\,.
\end{equation}
The resulting black holes then have the mass and $(n-1)$ independent charges.  It is now simply a matter of converting the $\xi$ coordinate to $r$ according to (\ref{xitor}) and selecting some appropriate parametrization for charges, since the original parameters are not conducive in describing the black hole solution. The procedure was outlined in detail with the $SL(4,\R)$ example in the previous section and we shall not repeat here.  The metric of the general $S(n,\R)$-Toda black hole is
\begin{eqnarray}
ds_D^2 &=& - U^{-1} f dt^2 + U^{\fft1{D-3}} (f^{-1} dt^2 + r^2 d\Omega_{D-2}^2)\,,\cr
U&=& (H_1 \cdots H_{n-1})^{\fft{12}{n(n^2-1)}}\,,\qquad
f=1 - \fft{m}{r^{D-3}}\,,
\end{eqnarray}
The dilaton scalars are given by
\begin{eqnarray}
&&e^{\vec a_1 \cdot \vec \phi} = \fft{H_1^2}{H_2} U^{-1}\,,\qquad
e^{\vec a_{n-1} \cdot \vec \phi} = \fft{H_{n-1}^2}{H_{n-2}} U^{-1}\,,\cr
&&e^{\vec a_i \cdot \vec \phi} = \fft{H_i^2}{H_{i-1} H_{i+1}} U^{-1}\,,\qquad \hbox{for}\qquad i=2,3,\ldots, n-2\,.
\end{eqnarray}
An equivalent expression for $\vec \phi$ is given by
\begin{equation}
\vec \phi= \ft{6(D-2)}{n(n^2-1)(D-3)} \sum_{i=1}^{n-1} \vec a_i \log H_i\,.
\end{equation}
The $U(1)$ fields are best given in terms of their field strengths:
\begin{eqnarray}
F_1 &=& \sqrt{\ft{12(D-2)(n-1)\,\beta_1\gamma_2}{(D-3)n(n^2-1)}} \,\fft{H_2}{\gamma_1 H_1^2}\, df\wedge dt\,,\cr
F_{n-1} &=& \sqrt{\ft{12(D-2)(n-1)\,\beta_{n-1} \gamma_{n-2}}{(D-3)n(n^2-1)}} \, \fft{H_{n-2}}{\gamma_{n-1} H_{n-1}^{2}}\, df\wedge dt\,,\cr
F_i &=& \sqrt{\ft{12(D-2)(n-1)\, \beta_i\gamma_{i-1}\gamma_{i+2} }{(D-3) n(n^2-1)}}\, \fft{H_{i-1}H_{i+1}}{\gamma_iH_i^{2}}\, df\wedge dt\,,
\quad i=2,3,\ldots,n-2\,.
\end{eqnarray}
The functions $H_i$ are given by $H_i={\cal H}_i(f)/\gamma_i$, and $\gamma_i={\cal H}_i(f=1)$ are constants.  The functions ${\cal H}_i$ are polynomials of $f$ of degrees $i(n-i)$.  To present these functions, we introduce $(n-1)$ numbers of independent parameters $\beta_i$, $i=1,2,\ldots, (n-1)$ and define
\begin{eqnarray}
&&c_{i+1} = - \ft{n-i}{i} \beta_i c_{i}\,,\qquad i=1,2,\ldots, (n-1)\,,\qquad \hbox{with}\qquad c_1=1\,,\cr
&&d_{k_1k_2\ldots k_i}= m^i c_{k_1} c_{k_2} \cdots c_{k_i} \prod_{k_1<\cdots<k_i} (k_i-k_j)^2\,.
\end{eqnarray}
We find
\begin{equation}
{\cal H}_i = \sum_{k_1<k_2<\cdots<k_i}^n \fft{d_{k_1k_2\cdots k_i}}{d_{12\cdots i}} f^{k_1 + k_2 + \cdots + k_i - \fft12i(i+1)}\,.
\end{equation}
When all the $\beta_i$ are equal, we have
\begin{equation}
{\cal H}_i = (1-\beta f)^{i(n-i)}\,,\qquad \hbox{for}\qquad \beta_i=\beta\,,
\end{equation}
and hence the solution becomes the RN black hole.  In general we have
\begin{equation}
{\cal H}_i = 1 - i(n-i) \beta_i f + \cdots\,,
\end{equation}
with the coefficient of the highest order of $f$ for each ${\cal H}_i$ given by
\begin{eqnarray}
{\cal H}_1 &=& 1 + \cdots + \beta_1 \cdots \beta_{n-1} f^{n-1}\,,\cr
{\cal H}_2 &=& 1 + \cdots + \beta_1 (\beta_2\cdots \beta_{n-2})^2 \beta_{n-1} f^{2(n-2)}\,,\cr
{\cal H}_3 &=& 1 + \cdots + \beta_1\beta_2^2 (\beta_3\cdots \beta_{n-3})^3
\beta_{n-2}^2\beta_{n-1} f^{3(n-3)}\,,\cr
&\vdots&\label{highestf}
\end{eqnarray}
Having obtained the explicit solution for general $SL(n,\R)$, let us give the thermodynamical quantities.  The mass is given by
\begin{equation}
M= \fft{(D-2)\Omega m}{16\pi} \fft{\partial (-g_{tt})}{\partial f} \Big|_{f\rightarrow 0}\,.
\end{equation}
The electric charges are given by
\begin{eqnarray}
Q_i &=& \ft{\Omega m}{16\pi \gamma_i} \sqrt{\ft{12i(n-i)}{n(n^2-1)} (D-2)(D-3)\beta_i \gamma_{i-1} \gamma_{i+1}}\,,\qquad i=1,2,\ldots,(n-1)\,.
\end{eqnarray}
with $\gamma_{0}\equiv 1\equiv \gamma_{n}$.  The horizon is located at $r_0=m^{1/(D-3)}$ and the temperature and the entropy is given by
\begin{equation}
T=\ft{D-3}{4\pi r_0} (\prod\gamma_i)^{\fft{6(D-2)}{n(n^2-1)(D-3)}}\,, \qquad
S=\ft14 \Omega \Big(\fft{m^{\fft16n(n^2-1)}}{\prod\gamma_i} \Big)^{ \fft{6(D-2)}{n(n^2-1)(D-3)}}\,.
\end{equation}
The electric potential differences between the horizon and the asymptotic infinite can be easily obtained, but the close-form expressions are lacking.  With many explicit examples, we find that the first law of thermodynamics is held, namely
\begin{equation}
dM = T dS + \sum_{i=1}^{n-1}\Phi_i dQ_i\,.
\end{equation}
In our parametrization, the free-energy is independent of the $\beta_i$'s:
\begin{equation}
F=M - T S - \sum_{i=1}^{n-1} \Phi_i Q_i = - \fft{\Omega\,m}{16\pi}\,.
\end{equation}
The inner horizon is located at $r=0$ for which the highest order of $f$ in each ${\cal H}_i$ dominates.  It follows from (\ref{highestf}) that the corresponding entropy
\begin{equation}
S_-=\ft14 \Big(m^{\fft16n(n^2-1)} \prod_{i=1}^{n-1} \gamma_i^{-1} \beta_i^{i(n-i)}\Big)^{ \fft{6(D-2)}{n(n^2-1)(D-3)}}\,.
\end{equation}
Thus we find that the product of the entropies of the outer and inner horizons,
\begin{equation}
S_+ S_- \sim \Big(\prod_i^{n-1} Q_i^{i(n-i)}\Big)^{ \fft{6(D-2)}{n(n^2-1)(D-3)}}\,,
\end{equation}
is expressed in terms of (quantized) charges only.  This is strongly suggestive of microscopic interpretation of the entropies in terms of some two-dimensional conformal field theories \cite{CYII,L,CLI,CLII}.

The exact expressions for extremal limit are complicated and we have not obtained the close form solutions for general $n$.  As in the case of $(D-3)$-branes of axion fields \cite{Lu:1996jr}, the function $H_i$ in our case are also polynomials in $\rho$ of degree $i(n-i)$.  For given $n$, the solution can be easily obtained.  The mass/charge relations are complicated and we shall not investigate further in this paper.

\section{Conclusions and discussions}

    In this paper, we constructed a minimum gravity theory in general
dimensions that would give rise to charged black hole solutions whose equations of motion are governed by a set of one-dimensional $SL(n,\R)$-Toda equations.  The theory consists of the metric, $(n-1)$ $U(1)$ vectors and $(n-2)$ dilatonic scalars.  The most general solution contains $2(n-1)$ non-trivial integration constants.  After eliminating $(n-2)$ scalar charges that either cause naked singularity or destroy the flat asymptotics, we obtained explicit black holes with the mass and $(n-1)$ independent charges. The $n=2$ case is simply the RN black hole and the $n=3$ example includes the well-known Kaluza-Klein dyon.  All the solutions with $n\ge 4$ are beyond supergravities.  The near-horizon geometries in the extremal limit are all AdS$_2\times S^{D-2}$. We also studied the thermodynamics and obtained the universal entropy product formula.  The entropy product formula implies that our black holes are of those with ``maximal'' charges, as defined in \cite{Cvetic:2013eda}. Indeed, for some appropriate charge configurations, {\it i.e.~}$\beta_i=\beta$ for all $i$, the solutions all become the RN black hole. These properties are suggestive of microscopic interpretation of the entropies using some two-dimensional conformal field theories, even though our theories in general cannot be embedded in supergravities.

     The black holes with $n\ge 3$ can all be viewed as bound states with negative binding
energy.  In the extremal limit, the energy of the black holes with $(n-1)$ independent charges is greater than the total sum of energy of each black holes associated with an individual charge.  Black holes can also come as bound states with positive and zero binding energies.  For a system with two independent charges, the mass/charge relation of known analytical black holes includes
\begin{eqnarray}
\hbox{Positive binding energy}:&& M \sim \sqrt{Q_1^2 + Q_2^2}\,,\cr
\hbox{Zero binding energy}:&& M \sim Q_1 + Q_2\,,\cr
\hbox{Negative binding energy}:&& M \sim \Big(Q_1^{\fft23} + Q_2^{\fft23}\Big)^{\fft32}\,,
\end{eqnarray}
The first type can arise in supergravities. For example, one can start with a singly-charged solution and then perform $SL(2,\R)$ U-duality group on the solution and obtain the two-charge solution with $(Q_1,Q_2)$ that form a doublet of the $SL(2,\R)$. Examples include the $(p,q)$ strings of type IIB supergravity \cite{Schwarz:1995dk} or any Kaluza-Klein black hole with two Kluza-Klein charges.  These solutions are beyond what we have discussed in this paper since they all involve additional axion fields that are associated with the scalar coset of U-duality groups. In fact there is yet an example of analytic solution of multi-charge black holes with positive binding energy of the Lagrangian (\ref{genlag}) without involving an axion field.  The second type is ubiquitous in supergravities, and the linear relation implies the balance of the gravitational and Maxwell forces, and hence multi-centered solutions characterized by some harmonic functions on the flat transverse space can be constructed.  Both the first and second types can be supersymmetric in supergravities.  The third type is the $SL(3,\R)$-Toda black holes including the Kaluza-Klein dyon.  We have generalized these to the $SL(n,\R)$-Toda black holes in this paper.

     The binding energy of the black holes from the Lagrangian (\ref{genlag}) is dictated
by the dot product $\vec a_i\cdot \vec a_j$.  Let
\begin{equation}
\vec a_i\cdot \vec a_j = - \fft{2(D-2)}{D-3}\gamma\,,\qquad i\ne j\,.
\end{equation}
Following the discussion in \cite{Lu:1996mv}, we find that for $0<\gamma<1$, the black hole has positive bind energy.  If $\gamma=1$, the black hole has zero binding energy.  For $\gamma>1$, the binding energy becomes negative.  For the $SL(n,\R)$ system in (\ref{aiajtoda}), we find that the adjacent ingredients have negative binding energy whilst non-adjacent ingredients have zero binding energy.

     Our work provides a large class of exact solutions of charged black hole with negative
binding energy.  Many intriguing properties of these black holes deserve further investigation.

\section*{Acknowledgement}

H.L.~is grateful to ICTS-USTC for hospitality during the course of this work.  The research is supported in part by the NSFC grants 11175269 and 11235003.

\end{document}